\documentclass[aps,pra,preprint,groupedaddress]{revtex4-1}
\usepackage{amsmath}
\usepackage{graphicx}
\begin{document}

%\preprint{}

\title{Deriving photon spin from relativistic quantum equations: Nonlocality of photon spin and relativistic quantum constraint}


\author{Chun-Fang Li}
\email[]{cfli@shu.edu.cn}
%\thanks{}
\affiliation{Department of Physics, Shanghai University, 99 Shangda Road, 200444 Shanghai, China}


\date{\today}

\begin{abstract}
The difficulties encountered up till now in the theory of identifying the spin and orbital angular momentum of the photon stem from the approach of dividing the angular momentum of the photon into spin and orbital parts.
Here we derive the spin of the photon from a set of two relativistic quantum equations that was first cast from the free-space Maxwell equations by Darwin in 1932. Much attention is focused on the nonlocal properties of the photon spin that are determined by the relativistic quantum constraint, one of the so-called Darwin equations.
Meanwhile, we point out that for the Darwin equations to describe the quantum motion of the photon, the upper and lower parts of the wavefunction cannot be the electric and magnetic fields as Darwin prescribed. Their nonlocal relations are investigated.
The Lorentz covariance of the Darwin equations is also proven, to the best of our knowledge, for the first time.
\end{abstract}

%\keywords{}


\maketitle


\newpage

\section{Introduction}

Experimental data revealed that the photon can have both spin and orbital angular momentum \cite{Fran-AP, Yao-P}. They can be distinguished by their different effects on tiny birefringent particles \cite{O'Neil-MAP, Garc-MPD}. They can also be converted from one into another \cite{Marr-MP, Zhao-EJM, Mosca, Marr}.
Nevertheless, a theoretical identification of the spin and orbital angular momentum of the photon has been a big challenge at length \cite{Darw, Akhi-B, Cohen-DG, van-N1, van-N2, Barn-A, Barn02, Li09, Li16E} and remains a hot topic of intense controversy.
Recently, much attention \cite{Barn10, Came, Blio-N12, Blio-DN, Bial, Barn14, Barn-AC, Lead, Lead18} was paid to the physical reality of the local density for the photon spin.
In particular, Bliokh \textit{et al} \cite{Blio-DN, Blio-N15, Blio-BN13, Blio-BN14, Blio-KN} still considered the vector matrix
$
\mathbf{\Omega}=\bigg(\begin{array}{cc}
                        \mathbf{\Sigma} & 0 \\
                        0               & \mathbf{\Sigma}
                      \end{array}
                \bigg)
$,
where $(\Sigma_k)_{ij} =-i \epsilon_{ijk}$ with $\epsilon_{ijk}$ the Levi-Civit\'{a} pseudotensor, as the operator for the photon spin to define its local density, even though van Enk and Nienhuis \cite{van-N1, van-N2} had shown that the operator for the photon spin in momentum representation has commuting components.

The difficulties encountered in theory stem from the approach of dividing the angular momentum of the photon into spin and orbital parts as Barnett pointed out in Ref. \cite{Barn10}.
The purpose of this paper is to put forward a new approach to identify the spin of the photon.
We will derive the photon spin from a set of relativistic quantum equations that was first advanced by Darwin \cite{Darw} and reformulated by Pryce \cite{Pryce} and Inagaki \cite{Inag}, paying particular attention to how the relativistic quantum constraint determines the nonlocality of the photon spin \cite{Bial}.

Darwin \cite{Darw} once showed that Maxwell's equations for a free radiation field can be cast into two equations about a six-component wavefunction, called Darwin's equations. One is a time-dependent Dirac-like equation \cite{Barn14}. The other has nothing to do with the time. It shows up as a relativistic quantum constraint (RQC) on the wavefunction since without it the Dirac-like equation cannot be relativistically covariant.
We find, however, that for Darwin's equations to describe the quantum motion of the photon, the upper and lower parts of the wavefunction cannot be the electric and magnetic fields as Darwin prescribed. They have a complicated nonlocal relation with the classical fields.
Considering Darwin's equations as quantum equations for the photon this way, we show that the same as the Dirac equation predicts the existence of the spin of the electron \cite{Dira}, the Dirac-like equation itself predicts the existence of a kind of spin. In particular, the spin appears as an intrinsic degree of freedom and is represented by the above-mentioned vector matrix $\mathbf \Omega$.
But when the role of the RQC is taken into account, the spin becomes non-intrinsic \cite{Bial-B, Li-Zh}. Its representative operator changes into
$(\mathbf{\Omega} \cdot \mathbf{p}) \mathbf{p}/p^2$
with $\mathbf p$ being the momentum operator, which has commuting components.
As a result, the so-called local density for the photon spin in position space is physically meaningless \cite{Bial}.
On the other hand, due to the RQC, the expectation value of the photon spin can be expressed in terms of the position-representation wavefunction as integrals of different integrands over the whole position space.
The key point is that the RQC conveys the nonlocality of the photon \cite{Jauch-P, Amre, Pauli, Rose-S} in such a way that the integral of the modulus squared of the position-representation wavefunction over the whole position space gives the total probability but the wavefunction itself does not mean the probability amplitude for the position of the photon.
All these results make up the main content of this paper. Let us first introduce the Darwin equations and briefly review their relationship with the Maxwell equations as well as their relativistic covariance.

\section{Darwin's Equations and Relativistic Quantum Constraint}\label{DEs}

According to Darwin \cite{Darw}, the relativistic quantum-mechanical description of free photons is done by a pair of equations about a six-component wavefunction $\Psi(\mathbf{x},t)$,
\begin{subequations}\label{REoM}
\begin{align}
  i \hbar \frac{\partial \Psi}{\partial t}  & =H \Psi,   \label{DLE} \\
  (\mathbf{\Gamma} \cdot \mathbf{p})^2 \Psi & =p^2 \Psi, \label{RQC}
\end{align}
\end{subequations}
where
$H=ic \Gamma_0 (\mathbf{\Gamma} \cdot \mathbf{p})$ is the Hamiltonian,
$c=1/(\varepsilon_0 \mu_0)^{1/2}$ is the speed of light in vacuum,
\begin{equation*}
    \Gamma_0=\bigg(
                   \begin{array}{cc}
                     I_3 & 0 \\
                     0   & -I_3
                   \end{array}
             \bigg),                \quad
    \mathbf{\Gamma}=\bigg(
                          \begin{array}{cc}
                            0               & \mathbf{\Sigma} \\
                            \mathbf{\Sigma} & 0
                          \end{array}
                    \bigg),
\end{equation*}
$I_3$ is the 3-by-3 unit matrix, and
$\mathbf{p}=-i \hbar \nabla$
is the momentum operator.
The matrices $\Gamma_0$ and $\mathbf{\Gamma}$ are all Hermitian and have the following properties,
\begin{subequations}
\begin{align}
  \Gamma^2_0 & = 1,                                           \label{p1} \\
  \Gamma_0 \mathbf{\Gamma} +\mathbf{\Gamma} \Gamma_0 & =0,    \label{p2} \\
  \Gamma_i \Gamma_j \Gamma_k +\Gamma_k \Gamma_j \Gamma_i &
 =\Gamma_i \delta_{jk} +\Gamma_k \delta_{ij}, \quad  i=1,2,3. \label{p3}
\end{align}
\end{subequations}
Eq. (\ref{DLE}) is a Dirac-like equation \cite{Barn14}. Here there is one peculiar feature that does not usually occur in quantum mechanics. This is that apart from satisfying the equation of motion (\ref{DLE}), the wavefunction has to satisfy Eq. (\ref{RQC}). From this equation it follows that
$H^2 =c^2 p^2$,
indicating that a photon with nonzero energy \cite{Darw} cannot have vanishing momentum.

It is easy to cast Darwin's equations (\ref{REoM}) into the form of Maxwell's equations in free space,
\begin{subequations}\label{ME}
\begin{align}
    \varepsilon_0 \frac{\partial \mathcal{E}}{\partial t}
   =\nabla \times \mathcal{H},  \quad &
    \mu_0 \frac{\partial \mathcal{H}}{\partial t}
   =-\nabla \times \mathcal{E},                           \label{ME1} \\
    \nabla \cdot \mathcal{E}=0, \quad &
    \nabla \cdot \mathcal{H}=0,                           \label{ME2}
\end{align}
\end{subequations}
where $\mathcal E$ and $\mathcal H$ are the real-valued electric and magnetic fields, respectively. In fact, letting
$\Psi=\frac{1}{\sqrt 2} \bigg(
                              \begin{array}{c}
                                \mathbf{F}_u \\
                                \mathbf{F}_l
                              \end{array}
                        \bigg)$
in accordance with the concrete forms of the matrices $\Gamma_0$ and $\mathbf \Gamma$, where the factor $\frac{1}{\sqrt 2}$ is introduced for later convenience, one readily changes Eq. (\ref{DLE}) into
\begin{equation*}
    i \hbar \frac{\partial \mathbf{F}_u}{\partial t}=-c \mathbf{p} \times \mathbf{F}_l, \quad
    i \hbar \frac{\partial \mathbf{F}_l}{\partial t}= c \mathbf{p} \times \mathbf{F}_u,
\end{equation*}
or, equivalently,
\begin{equation}\label{CEoM}
    \frac{\partial \mathbf{F}_u}{\partial t}= c \nabla \times \mathbf{F}_l, \quad
    \frac{\partial \mathbf{F}_l}{\partial t}=-c \nabla \times \mathbf{F}_u,
\end{equation}
where the relation
$(\mathbf{\Sigma} \cdot \mathbf{a}) \mathbf{b}= i \mathbf{a} \times \mathbf{b}$
has been used. They have the same form as Maxwell's equations (\ref{ME1}) if the following correspondences are assumed,
\begin{equation*}
    \mathbf{F}_u \sim \sqrt{\varepsilon_0} \mathcal{E}, \quad
    \mathbf{F}_l \sim \sqrt{\mu_0}         \mathcal{H}.
\end{equation*}
Meanwhile, Eq. (\ref{RQC}) can be rewritten in terms of $\mathbf{F}_u$ and $\mathbf{F}_l$ as
\begin{equation}\label{RCC}
    \mathbf{p} (\mathbf{p} \cdot \mathbf{F}_u)=0, \quad
    \mathbf{p} (\mathbf{p} \cdot \mathbf{F}_l)=0,
\end{equation}
or, equivalently,
\begin{equation}\label{RCC-r}
    \nabla (\nabla \cdot \mathbf{F}_u)=0, \quad
    \nabla (\nabla \cdot \mathbf{F}_l)=0.
\end{equation}
When the energy does not vanish, the momentum does not, either, $\mathbf{p} \neq 0$. In this case, Eqs. (\ref{RCC}) mean
$\mathbf{p} \cdot \mathbf{F}_u =0$ and $\mathbf{p} \cdot \mathbf{F}_l =0$.
That is to say,
\begin{equation}\label{EoD}
    \nabla \cdot \mathbf{F}_u =0, \quad \nabla \cdot \mathbf{F}_l =0.
\end{equation}
They are the same as Maxwell's equations (\ref{ME2}). Moreover, when the energy vanishes, that is to say, when the wavefunction satisfies
$H \Psi =0$, one has
$\frac{\partial \Psi}{\partial t} =0$
in accordance with Eq. (\ref{DLE}).
In such a case, Eqs. (\ref{CEoM}) become
\begin{equation*}
    \nabla \times \mathbf{F}_u =0, \quad \nabla \times \mathbf{F}_l =0,
\end{equation*}
which, together with Eqs. (\ref{RCC-r}), lead to
\begin{equation}\label{LE}
    \nabla^2 \mathbf{F}_u =0, \quad \nabla^2 \mathbf{F}_l =0.
\end{equation}
According to Stratton \cite{Stra}, the vector functions that throughout all space satisfy Laplace's equations (\ref{LE}) vanish at infinity,
$\mathbf{F}_u |_{\infty} =\mathbf{F}_l |_{\infty} =0$. From the uniqueness theorem \cite{Jack} it follows that Eqs. (\ref{LE}) with these boundary conditions have only trivial solutions,
$\mathbf{F}_u =0$ and $\mathbf{F}_l =0$,
which satisfy Eqs. (\ref{EoD}). In a word, Darwin's equations (\ref{REoM}) can indeed be cast into the form of Maxwell's equations in free space.
However, it should be pointed out that the upper and lower parts of the wavefunction $\Psi$ cannot be the electric and magnetic fields as Darwin \cite{Darw} prescribed. An evident fact is that the wavefunction is not necessarily real-valued the same as the electric and magnetic fields. We will come back to this problem at the end of Section \ref{PPS}.

From the above discussions one can infer that Darwin's equations (\ref{REoM}) are Lorentz covariant.
Indeed, letting
$(ict,x,y,z) \equiv (x_0,x_1,x_2,x_3)$, Eq. (\ref{DLE}) can be rewritten as
\begin{equation*}
    \Gamma_\mu p_\mu \Psi=0, \quad \mu=0,1,2,3,
\end{equation*}
where
$p_0=-i \hbar \frac{\partial}{\partial x_0}$
and summation convention has been used. Multiplying this equation by $\Gamma_\nu p_\nu$ on the left and using Eqs. (\ref{p1}) and (\ref{p2}), we have
\begin{equation*}
    [p^2_0 +(\mathbf{\Gamma} \cdot \mathbf{p})^2] \Psi=0.
\end{equation*}
It is of course not Lorentz covariant. But upon substituting Eq. (\ref{RQC}), we get the following Klein-Gordon equation for zero-mass particles \cite{Darw},
\begin{equation*}
    p_\mu p_\mu \Psi=0,
\end{equation*}
which is Lorentz covariant. This shows that Eq. (\ref{RQC}) is necessary for the Dirac-like equation (\ref{DLE}) to be Lorentz covariant.
In other words, Eq. (\ref{RQC}) acts as a RQC on the photon.
The transformation formula for the wavefunction under Lorentz transformations is given in Appendix \ref{LT} and the Lorentz covariance of Darwin's equations (\ref{REoM}) is shown in Appendix \ref{LC}.
We will see in the next section that the Dirac-like equation itself implies the existence of the spin.

\section{Spin Predicted Solely by Dirac-like Equation}\label{CanoS}

Putting the RQC (\ref{RQC}) aside, it is not difficult to show by use of the Dirac-like equation (\ref{DLE}) that the orbital angular momentum
$\mathbf{L}=\mathbf{x} \times \mathbf{p}$
is not a constant of motion,
\begin{equation*}
    [H, \mathbf{L}]=\hbar c \Gamma_0 \mathbf{\Gamma} \times \mathbf{p},
\end{equation*}
where we have assumed the following commutation relations,
\begin{equation}\label{CR-PandX}
    [p_i, p_j] =0, \quad [x_i, p_j] =i \hbar \delta_{ij}.
\end{equation}
For this reason, we introduce the constant vector matrix
$-i \mathbf{\Gamma} \times \mathbf{\Gamma}$, which is exactly equal to
$\mathbf{\Omega}$ according to the commutation relation
\begin{equation}\label{CR-Sigma}
    [\Sigma_i, \Sigma_j]=i \epsilon_{ijk} \Sigma_k.
\end{equation}
With the help of Eqs. (\ref{p2}) and (\ref{p3}), it is easy to find
\begin{equation*}
    [H, \mathbf{\Omega}]=-c \Gamma_0 \mathbf{\Gamma} \times \mathbf{p}.
\end{equation*}
So we see that the sum of $\mathbf L$ and $\hbar \mathbf{\Omega}$ is a constant of motion.
Because it is independent of the extrinsic degrees of freedom such as the position and momentum, $\mathbf \Omega$ represents an intrinsic degree of freedom, called the spin. It obeys the canonical commutation relation,
\begin{equation}\label{CR-Omega}
    [\Omega_i, \Omega_j]=i \epsilon_{ijk} \Omega_k,
\end{equation}
by virtue of Eq. (\ref{CR-Sigma}). The constant of motion,
$\mathbf{L}+\hbar \mathbf{\Omega}$,
is the total angular momentum.
In a word, the spin is distinguishable from the orbital angular momentum as long as the momentum $\mathbf p$ is well defined.

According to the definition \cite{Saku}, the expectation value of the spin in an arbitrary state $\Psi$ that is normalized as
$\int \Psi^\dag \Psi d^3 x=1$
is given by
\begin{equation}\label{EoS}
    \langle\mathbf{\Omega}\rangle=\int \Psi^\dag \mathbf{\Omega} \Psi d^3 x
\end{equation}
in units of $\hbar$. Being an intrinsic degree of freedom, the spin here can also be represented in momentum representation by the same constant vector matrix $\mathbf \Omega$. In fact, denoting by $\psi(\mathbf{k},t)$ the wavefunction in momentum representation with $\mathbf k$ the wavevector, which is the Fourier component of $\Psi$,
\begin{equation}\label{FT}
    \Psi(\mathbf{x}, t)=\frac{1}{(2 \pi)^{3/2}}
    \int \psi(\mathbf{k}, t) \exp(i \mathbf{k} \cdot \mathbf{x}) d^3 k,
\end{equation}
we readily change Eq. (\ref{EoS}) into
\begin{equation*}
    \langle\mathbf{\Omega}\rangle=\int \psi^\dag \mathbf{\Omega} \psi d^3 k.
\end{equation*}
Nevertheless, it is noted that the constant operator $\mathbf \Omega$ does not represent the spin of the photon.
This is because the canonical commutation relation (\ref{CR-Omega}) together with the property
$\mathbf{\Omega}^2=2$
leads to a consequence \cite{Saku} that the component of $\mathbf \Omega$ in any fixed direction has eigenvalues of $\pm 1$ and $0$, which is apparently not the property of the photon spin.
Fortunately, the wavefunction of the photon has to obey, apart from the Dirac-like equation (\ref{DLE}), the RQC (\ref{RQC}), which has not yet been exploited at all. Observing that the occurrence of the RQC does not hinder the separation of the spin from the orbital angular momentum, let us examine how the RQC determines the properties of the photon spin.

\section{Nonlocality of Photon Spin Determined by RQC}\label{PPS}

\subsection{Nonexistence of local density for photon spin in position space}

\subsubsection{Darwin's equations in momentum representation}

Considering that the RQC (\ref{RQC}) is expressed in terms of the momentum, it is beneficial to write out the Darwin equations (\ref{REoM}) in momentum representation.
To this end, we substitute Eq. (\ref{FT}) to get
\begin{subequations}\label{REoM-k}
\begin{align}
  i \hbar \frac{\partial \psi}{\partial t}  & =H   \psi,     \label{DLE-k} \\
  (\mathbf{\Gamma} \cdot \mathbf{k})^2 \psi & =k^2 \psi,     \label{RQC-k}
\end{align}
\end{subequations}
where
$H=i \hbar c \Gamma_0 (\mathbf{\Gamma} \cdot \mathbf{k})$
and $k=|\mathbf{k}|$. From Eq. (\ref{RQC-k}) we have
$H^2 =\hbar^2 c^2 k^2$.
It shows that Eq. (\ref{DLE-k}) has solutions of negative as well as positive energies.
Akin to solutions to Dirac's equation \cite{Pesk-S}, solutions of negative energy correspond to antiphotons \cite{Bial96} if solutions of positive energy correspond to photons.
In this paper, we are not concerned with photon annihilation and creation and hence will consider only solutions of positive energy. Taking this into account, it is observed that $\psi$ must be the eigenfunction of the Hamiltonian with eigenvalue $\hbar \omega$, obeying the following eigenvalue equation,
\begin{equation}\label{EVE}
    ic \Gamma_0 (\mathbf{\Gamma} \cdot \mathbf{k}) \psi =\omega \psi,
\end{equation}
where $\omega=c k$. Substituting it into Eq. (\ref{DLE-k}) and
letting
$
\psi=\frac{1}{\sqrt 2} \bigg(
                             \begin{array}{c}
                               \mathbf{f}_u \\
                               \mathbf{f}_l
                             \end{array}
                       \bigg)
$,
where $\mathbf{f}_u$ and $\mathbf{f}_l$ are the Fourier transformations of $\mathbf{F}_u$ and $\mathbf{F}_l$, respectively,
\begin{equation}\label{FT-f}
    \mathbf{f}_{u,l}(\mathbf{k},t)=\frac{1}{(2 \pi)^{3/2}}
    \int \mathbf{F}_{u,l}(\mathbf{x},t) \exp(-i \mathbf{k} \cdot \mathbf{x}) d^3 x,
\end{equation}
we have
\begin{equation}\label{SE}
    i \frac{\partial \mathbf{f}_{u,l}}{\partial t}=\omega \mathbf{f}_{u,l}.
\end{equation}
It says that both $\mathbf{f}_u$ and $\mathbf{f}_l$ are the eigenfunctions of the energy operator
$E= i \hbar \frac{\partial}{\partial t}$,
having the same eigenvalue $\hbar \omega$.
At the same time, since no photon state can have $k=0$, Eq. (\ref{RQC-k}) can be rewritten in terms of $\mathbf{f}_u$ and $\mathbf{f}_l$ as
\begin{equation}\label{TC-k}
    \mathbf{k} \cdot \mathbf{f}_{u,l}=0,
\end{equation}
in agreement with the eigenvalue equation (\ref{EVE}), which reads
\begin{equation}\label{CE-k}
    \mathbf{k} \times \mathbf{f}_l=-k \mathbf{f}_u , \quad
    \mathbf{k} \times \mathbf{f}_u= k \mathbf{f}_l .
\end{equation}

The Darwin equations (\ref{REoM}) are thus converted into Eqs. (\ref{SE})-(\ref{CE-k}) in momentum representation under the condition of positive energy. Eq. (\ref{SE}) corresponds to the Dirac-like equation (\ref{DLE}). Eq. (\ref{TC-k}) or (\ref{CE-k}) corresponds to the RQC (\ref{RQC}).

\subsubsection{Photon spin operator in momentum representation}

Now we are in a position to discuss how the RQC (\ref{TC-k}) determines the nonlocal property of the photon spin. Upon taking Eq. (\ref{TC-k}) into consideration, we must have
\begin{equation*}
    \mathbf{f}^\dag_{u,l}(\mathbf{\Sigma} \times \mathbf{w})
    \mathbf{f}_{u,l}=0,
\end{equation*}
where
$\mathbf{a}^\dag \Sigma_i \mathbf{b}
=-i (\mathbf{a}^* \times \mathbf{b})_i$
and $\mathbf{w}=\mathbf{k}/k$ stands for the unit momentum. Using this property to decompose the vector matrix $\mathbf \Sigma$ as
\begin{equation*}
    \mathbf{\Sigma}=\mathbf{\Sigma} \cdot \mathbf{w} \mathbf{w}
               -(\mathbf{\Sigma} \times \mathbf{w}) \times \mathbf{w},
\end{equation*}
we find
\begin{equation}\label{redu-Si}
    \mathbf{f}^\dag_{u,l} \mathbf{\Sigma} \mathbf{f}_{u,l}
   =[\mathbf{f}^\dag_{u,l} (\mathbf{\Sigma} \cdot \mathbf{w}) \mathbf{f}_{u,l}]
     \mathbf{w},
\end{equation}
which is equivalent to
\begin{equation}\label{redu-Om}
    \psi^\dag \mathbf{\Omega} \psi
   =\psi^\dag (\mathbf{\Omega} \cdot \mathbf{w} \mathbf{w}) \psi,
\end{equation}
where $\mathbf{w} \mathbf{w}$ is a dyadic.
Based on the arbitrariness of $\psi$ it is concluded that the RQC in momentum representation changes the spin operator from $\mathbf \Omega$ to $\mathbf{\Omega} \cdot \mathbf{w} \mathbf{w}$.
That is to say, the operator for the photon spin in momentum representation is no longer the constant vector matrix $\mathbf \Omega$. It becomes
\begin{equation}
    \mathbf{S}=(\mathbf{\Omega} \cdot \mathbf{w}) \mathbf{w},
\end{equation}
which is dependent on the momentum.
In the first place, it coincides with the well-known conclusion that the spin of the photon is always oriented in its propagation direction \cite{Jauch-R}. In the second place, it has commuting Cartesian components,
\begin{equation*}
    [S_i, S_j]=0,
\end{equation*}
in perfect agreement with the result that was obtained in the framework of second quantization \cite{van-N1, van-N2}.
In the third place, commuting with the Hamiltonian
$[H, \mathbf{S}]=0$,
it is now a constant of motion.
In a word, the RQC makes the spin of the photon depend on the momentum of the photon. It is no longer an independent degree of freedom \cite{Bial-B}.
As a result, the expectation value of the spin of the photon in an arbitrary state $\psi$ is given by
\begin{equation}\label{EoPS}
    \langle \mathbf{S} \rangle
   =\int\psi^\dag (\mathbf{\Omega} \cdot \mathbf{w} \mathbf{w}) \psi d^3 k.
\end{equation}

The occurrence of the dyadic $\mathbf{ww}$ in $\mathbf{S}$ results in the absence of the local density for the photon spin in position space.
To see this, we substitute the inverse Fourier transformation of Eq. (\ref{FT}) into Eq. (\ref{EoPS}) to get
\begin{equation*}
    \langle \mathbf{S} \rangle=\int \mathbf{s}(\mathbf{x},t) d^3 x,
\end{equation*}
where
\begin{equation}\label{s}
    \mathbf{s}(\mathbf{x},t)
   =\Psi^\dag (\mathbf{x},t) \int \mathbf{G} (\mathbf{x}-\mathbf{x}')
                                        \Psi (\mathbf{x}',t) d^3 x',
\end{equation}
\begin{equation*}
    \mathbf{G} (\mathbf{x}) =\frac{1}{(2 \pi)^3} \mathbf{\Omega} \cdot
       \int \mathbf{w} \mathbf{w} \exp(i \mathbf{k} \cdot \mathbf{x}) d^3 k.
\end{equation*}
The integrand $\mathbf s$ does not locally depend on the wavefunction $\Psi$. Its value at any particular point $\mathbf x$ depends not only on the value of the wavefunction at that point but also on the value at all other points.
As a result, the integrand $\mathbf s$ cannot be interpreted as the local density for the photon spin in position space though its integral over the whole position space yields the expectation value. This shows the nonlocality of the photon spin in position space \cite{Bial}.
It is interesting to note that similar to $\mathbf s$ in (\ref{s}), the integrand of the expression (8) in Ref. \cite{Li16E} cannot be interpreted as the local density for the spin angular momentum of classical radiation fields.
This is because the expression (5) there cannot be converted into a function that locally depends on the electric and magnetic fields due to the occurrence of the unit momentum
$\mathbf w$ in the integrands.

If $\mathbf{w}\mathbf{w}$ is replaced with the unit dyadic, $\mathbf{G} (\mathbf{x})$ will be replaced with
$\mathbf{\Omega} \delta^3 (\mathbf{x})$
and $\mathbf{s}(\mathbf{x}, t)$ will be replaced with
$\Psi^\dag \mathbf{\Omega} \Psi$. In this case, $\mathbf \Omega$ will represent the spin in position representation.
Since $\mathbf{w}\mathbf{w}$ is not the unit dyadic, $\mathbf \Omega$ cannot be the operator for the spin of the photon in position representation as Bliokh \textit{et al} claimed \cite{Blio-DN, Blio-N15, Blio-BN13, Blio-BN14, Blio-KN}.
However, we will see in the following that the RQC allows to express the expectation value of the photon spin in terms of the position-representation wavefunction as integrals of different integrands over the whole position space.

\subsection{Further discussions on expectation value of photon spin}

\subsubsection{Expectation value of photon spin expressed in position representation}

We have shown that the RQC (\ref{TC-k}) in momentum representation leads to Eq. (\ref{redu-Om}), which allows to change Eq. (\ref{EoPS}) into
\begin{equation*}
    \langle \mathbf{S} \rangle
   =\int \psi^\dag \mathbf{\Omega} \psi d^3 k.
\end{equation*}
Upon substituting the inverse Fourier transformation of (\ref{FT}), we arrive at
\begin{equation*}
    \langle \mathbf{S} \rangle
   =\int \Psi^\dag \mathbf{\Omega} \Psi d^3 x,
\end{equation*}
which is the same as Eq. (\ref{EoS}). This is understandable. As we showed in Section \ref{CanoS}, the expectation value of the spin can always be expressed by Eq. (\ref{EoS}) as long as its wavefunction satisfies the Dirac-like equation (\ref{DLE}).
The result that $\mathbf \Omega$ is not the photon spin operator in position representation comes from the RQC (\ref{RQC}). It means that the integrand $\Psi^\dag \mathbf{\Omega} \Psi$ cannot be interpreted as the local density for the photon spin in position space.

To show this, we rewrite Eq. (\ref{EoPS}) in terms of the upper and lower parts of $\psi$ as
\begin{equation*}
    \langle \mathbf{S} \rangle =\frac{1}{2}
    \int[\mathbf{f}_u^\dag (\mathbf{\Sigma} \cdot \mathbf{w}) \mathbf{f}_u
        +\mathbf{f}_l^\dag (\mathbf{\Sigma} \cdot \mathbf{w}) \mathbf{f}_l] \mathbf{w} d^3 k.
\end{equation*}
Observing that
\begin{equation*}
    \mathbf{f}_u^\dag (\mathbf{\Sigma} \cdot \mathbf{w}) \mathbf{f}_u
   =\mathbf{f}_l^\dag (\mathbf{\Sigma} \cdot \mathbf{w}) \mathbf{f}_l,
\end{equation*}
by virtue of Eq. (\ref{CE-k}), we have
\begin{equation*}
    \langle \mathbf{S} \rangle
   =\int [\mathbf{f}^\dag_{u,l} (\mathbf{\Sigma} \cdot \mathbf{w}) \mathbf{f}_{u,l}]
          \mathbf{w} d^3 k.
\end{equation*}
Resorting to the property (\ref{redu-Si}), we get
\begin{equation}\label{EoS-k}
    \langle \mathbf{S} \rangle
   =\int \mathbf{f}^\dag_{u,l} \mathbf{\Sigma} \mathbf{f}_{u,l} d^3 k
   =-i \int \mathbf{f}^*_{u,l} \times \mathbf{f}_{u,l} d^3 k.
\end{equation}
Upon substituting the Fourier transformation (\ref{FT-f}), we find
\begin{equation}\label{EoS-F}
    \langle \mathbf{S} \rangle
   =-i \int \mathbf{F}^*_{u,l} \times \mathbf{F}_{u,l} d^3 x,
\end{equation}
which can be further expressed in terms of the position-representation wavefunction $\Psi$ as
\begin{equation*}
    \langle \mathbf{S} \rangle
   =\int \Psi^\dag (1 \pm \Gamma_0) \mathbf{\Omega} \Psi d^3 x.
\end{equation*}
Generally speaking, neither
$\Psi^\dag (1 +\Gamma_0) \mathbf{\Omega} \Psi$
nor
$\Psi^\dag (1 -\Gamma_0) \mathbf{\Omega} \Psi$
is equal to
$\Psi^\dag \mathbf{\Omega} \Psi$.
Depending locally on the wavefunction $\Psi$, they are also not equal to $\mathbf{s} (\mathbf{x},t)$ in Eq. (\ref{s}). It is thus concluded that there is no unique local density for the photon spin in position space \cite{Barn10, Barn14}.
This is what we mean here by the nonlocality of the photon spin in position space.

It is important to note that the expectation value of the orbital angular momentum, defined by
$\langle \mathbf{L} \rangle=\int \Psi^\dag \mathbf{L} \Psi d^3 x$,
can be converted in the same way into
\begin{equation}\label{EoL-k}
    \langle \mathbf{L} \rangle
   =-i \hbar \int \mathbf{f}_{u,l}^\dag
                 (\mathbf{k} \times \nabla_{\mathbf k}) \mathbf{f}_{u,l} d^3 k
\end{equation}
in momentum representation or into
\begin{equation}\label{EoL}
    \langle \mathbf{L} \rangle
   =-i \hbar \int \mathbf{F}_{u,l}^\dag
                 (\mathbf{x} \times \nabla) \mathbf{F}_{u,l} d^3 x
\end{equation}
in position representation, where $\nabla_{\mathbf k}$ denotes the gradient operator with respect to $\mathbf k$.

Eqs. (\ref{EoS-F}) and (\ref{EoL}) show that the expectation values of both the spin and orbital angular momentum can be expressed in terms solely of the upper or the lower part of the wavefunction. Let us analyze how the upper and lower parts of the wavefunction are related to the electric and magnetic fields of a free radiation field.

\subsubsection{Nonlocal relation of quantum wavefunction with classical fields}

It is well known that the real-valued electric and magnetic fields of a free radiation field, when expressed as \cite{Akhi-B}
\begin{equation}\label{EH-R}
    \mathcal{E}=\frac{1}{\sqrt 2}(\mathbf{E} +\mathbf{E}^*), \quad
    \mathcal{H}=\frac{1}{\sqrt 2}(\mathbf{H} +\mathbf{H}^*),
\end{equation}
can be expanded in terms of the plane-wave modes in the following way,
\begin{subequations}\label{EH}
\begin{align}
  \mathbf{E} (\mathbf{x},t) &= \frac{1}{(2 \pi)^{3/2}}
 \int\mathbf{e}(\mathbf{k},t)\exp(i\mathbf{k} \cdot \mathbf{x}) d^3 k,\label{CE} \\
  \mathbf{H} (\mathbf{x},t) &= \frac{1}{(2 \pi)^{3/2}}
 \int \mathbf{h}(\mathbf{k},t) \exp(i \mathbf{k} \cdot \mathbf{x}) d^3 k,
\end{align}
\end{subequations}
where the expansion coefficients $\mathbf e$ and $\mathbf h$ are related to each other via
\begin{equation*}
    \mathbf{h}= \frac{1}{\mu_0 c} \mathbf{w} \times \mathbf{e}, \quad
    \mathbf{e}=-\frac{1}{\varepsilon_0 c} \mathbf{w} \times \mathbf{h}
\end{equation*}
by virtue of Maxwell's equations (\ref{ME}).
In terms of the expansion coefficients, the spin and orbital angular momentum identified in classical theory \cite{Li09, Li16E} are expressed as
\begin{eqnarray*}
  -i \int \frac{\varepsilon_0}{ck} \mathbf{e}^* \times \mathbf{e} d^3 k
   \quad & \mathrm{or} \quad &
     -i \int \frac{\mu_0}{ck}         \mathbf{h}^* \times \mathbf{h} d^3 k, \\
  -i \int \frac{\varepsilon_0}{ck} \mathbf{e}^*
             (\mathbf{k} \times \nabla_{\mathbf k})       \mathbf{e} d^3 k
   \quad & \mathrm{or} \quad &
  -i \int \frac{\mu_0}{ck}         \mathbf{h}^*
             (\mathbf{k} \times \nabla_{\mathbf k})       \mathbf{h} d^3 k,
\end{eqnarray*}
respectively.
If they are postulated to be equal to their corresponding expectation values in quantum mechanics, a comparison with Eqs. (\ref{EoS-k}) and (\ref{EoL-k}) leads to the following relations,
\begin{equation}\label{CR}
    \mathbf{f}_u =\Big( \frac{\varepsilon_0}{\hbar c k} \Big)^{1/2}
                  \mathbf{e}, \quad
    \mathbf{f}_l =\Big( \frac{\mu_0}{\hbar c k} \Big)^{1/2} \mathbf{h}.
\end{equation}
With these relations, we are ready to show how the quantum wavefunction $\Psi$ in position representation is related to the classical fields.

First of all, the quantum wavefunction should be complex-valued.
In addition, since $\frac{1}{\sqrt k}$ is the Fourier transformation of
$\frac{1}{2 |\mathbf{x}|^{5/2}}$,
\begin{equation*}
    \frac{1}{\sqrt k}
   =\frac{1}{(2 \pi)^{3/2}}
    \int \frac{1}{2 |\mathbf{x}|^{5/2}} \exp(-i \mathbf{k} \cdot \mathbf{x}) d^3 x,
\end{equation*}
it follows from Eqs. (\ref{CR}), (\ref{FT}), and (\ref{EH}) that the upper and lower parts of the wavefunction are expressed in terms of the complex vector functions $\mathbf E$ and $\mathbf H$ as \cite{Cook}
\begin{subequations}\label{Psi}
\begin{align}
   \mathbf{F}_u (\mathbf{x},t) &
  =\sqrt{\frac{\varepsilon_0}{2 \pi \hbar c}} \int
   \frac{\mathbf{E}(\mathbf{x}',t)}{4 \pi |\mathbf{x}-\mathbf{x}'|^{5/2}} d^3 x',\\
   \mathbf{F}_l (\mathbf{x},t) &
  =\sqrt{\frac{\mu_0}{2 \pi \hbar c}} \int \frac{\mathbf{H}(\mathbf{x}',t)}
   {4 \pi |\mathbf{x}-\mathbf{x}'|^{5/2}} d^3 x',
\end{align}
\end{subequations}
respectively. The upper (or lower) part of the quantum wavefunction taken at one particular point $\mathbf x$ depends not only on the value of the classical function $\mathbf E$ (or $\mathbf H$) at that point but also on the value at all other points.
That is to say, the quantum wavefunction does not locally depend on the classical functions $\mathbf E$ and $\mathbf H$.
More importantly, the complex functions $\mathbf E$ and $\mathbf H$ do not locally depend on the real-valued electric and magnetic fields $\mathcal{E}$ and $\mathcal{H}$.

To show this, denoting by $\bm{\varepsilon} (\mathbf{k},t)$ and $\bm{\eta} (\mathbf{k},t)$ the Fourier transformations of $\mathcal{E}$ and $\mathcal{H}$, respectively,
\begin{subequations}
\begin{align}
  \mathcal{E} (\mathbf{x},t) &= \frac{1}{(2 \pi)^{3/2}} \int \bm{\varepsilon} (\mathbf{k},t) \exp(i \mathbf{k} \cdot \mathbf{x}) d^3 k, \label{RE} \\
  \mathcal{H} (\mathbf{x},t) &= \frac{1}{(2 \pi)^{3/2}} \int \bm{\eta}    (\mathbf{k},t) \exp(i \mathbf{k} \cdot \mathbf{x}) d^3 k, \label{RH}
\end{align}
\end{subequations}
a comparison with Eq. (\ref{EH-R}) gives
\begin{equation*}
    \bm{\varepsilon}(\mathbf{k},t)=\frac{1}{\sqrt 2}
    [\mathbf{e}(\mathbf{k},t)+\mathbf{e}^\ast (-\mathbf{k},t)], \quad
    \bm{\eta}(\mathbf{k},t)       =\frac{1}{\sqrt 2}
    [\mathbf{h}(\mathbf{k},t)+\mathbf{h}^\ast (-\mathbf{k},t)],
\end{equation*}
which have the properties,
\begin{equation}\label{symm}
    \bm{\varepsilon}^\ast (-\mathbf{k},t)=\bm{\varepsilon}(\mathbf{k},t), \quad
    \bm{\eta}^\ast (-\mathbf{k},t)=\bm{\eta}(\mathbf{k},t).
\end{equation}
It is clear that $\mathbf e$ and $\mathbf h$, the Fourier transformations of the complex functions $\mathbf E$ and $\mathbf H$, are different from $\bm{\varepsilon}$ and $\bm{\eta}$. They are not constrained by such conditions as Eqs. (\ref{symm}) \cite{Akhi-B}.
But they can be expressed in terms of $\bm{\varepsilon}$ and $\bm{\eta}$ as \cite{Cohen-DG}
\begin{subequations}
\begin{align}
  \mathbf{e}(\mathbf{k},t) & =\frac{1}{\sqrt 2}
    [\bm{\varepsilon}(\mathbf{k},t)-\frac{\mu_0 c}{k}
     \mathbf{k} \times \bm{\eta}(\mathbf{k},t)],        \label{e-epsi+eta}\\
  \mathbf{h}(\mathbf{k},t) & =\frac{1}{\sqrt 2}
    [\bm{\eta}(\mathbf{k},t)+\frac{\varepsilon_0 c}{k}
     \mathbf{k} \times \bm{\varepsilon}(\mathbf{k},t)], \label{h-eta+epsi}
\end{align}
\end{subequations}
by virtue of the Maxwell equations (\ref{ME}). According to Eq. (\ref{RH}) and the first Maxwell equation in (\ref{ME1}), we have
\begin{equation*}
    \frac{1}{(2 \pi)^{3/2}} \int (\mathbf{k} \times \bm{\eta})
    \exp(i \mathbf{k} \cdot \mathbf{x}) d^3 k
   =-i \varepsilon_0 \frac{\partial \mathcal{E}}{\partial t}.
\end{equation*}
Since $\frac{1}{k}$ is the Fourier transformation of
$\sqrt{\frac{2}{\pi}} \frac{1}{|\mathbf x|^2}$,
\begin{equation}\label{FI-1/k}
    \frac{1}{k}=\frac{1}{(2 \pi)^{3/2}}
    \int \sqrt{\frac{2}{\pi}} \frac{1}{|\mathbf x|^2}
    \exp(-i \mathbf{k} \cdot \mathbf{x}) d^3 x,
\end{equation}
it follows from Eqs. (\ref{e-epsi+eta}), (\ref{CE}), and (\ref{RE}) that the complex function $\mathbf E$ is related to the electric field $\mathcal{E}$ in the following way \cite{Pauli},
\begin{equation}\label{CE-RE}
    \mathbf{E}(\mathbf{x},t)=\frac{1}{\sqrt 2}
   \Big[\mathcal{E}(\mathbf{x},t)
        +\frac{i}{2 \pi^2 c} \frac{\partial}{\partial t}
         \int \frac{\mathcal{E}(\mathbf{x}',t)}{|\mathbf{x}-\mathbf{x}'|^2} d^3 x' \Big].
\end{equation}
The real part is proportional to the electric field $\mathcal E$ as the first equation in (\ref{EH-R}) requires. But the imaginary part does not locally depend on the electric field.
Similarly, the complex function $\mathbf H$ is related to the magnetic field $\mathcal H$ as follows,
\begin{equation}\label{CH-RH}
    \mathbf{H}(\mathbf{x},t)=\frac{1}{\sqrt 2}
   \Big[ \mathcal{H}(\mathbf{x},t)
        +\frac{i}{2 \pi^2 c} \frac{\partial}{\partial t}
         \int \frac{\mathcal{H}(\mathbf{x}',t)}{|\mathbf{x}-\mathbf{x}'|^2} d^3 x' \Big],
\end{equation}
where we have used Eq. (\ref{FI-1/k}) and the following relation,
\begin{equation*}
    \frac{1}{(2 \pi)^{3/2}} \int (\mathbf{k} \times \bm{\varepsilon})
    \exp(i \mathbf{k} \cdot \mathbf{x}) d^3 k
   =i \mu_0 \frac{\partial \mathcal{H}}{\partial t}.
\end{equation*}
In a word, the complex functions $\mathbf E$ and $\mathbf H$ do not locally depend on the electric and magnetic fields.

It is thus concluded from Eqs. (\ref{Psi}), (\ref{CE-RE}), and (\ref{CH-RH}) that the upper and lower parts of the wavefunction do not locally depend on the electric and magnetic fields.
Eqs. (\ref{Psi}) show that the wavefunction satisfying the Darwin equations is similar to the so-called Landau-Peierls wavefunction \cite{Land-P} that was discussed by Pauli \cite{Pauli} and Bialynicki-Birula \cite{Bial96},
except that the complex functions $\mathbf E$ and $\mathbf H$ in Eqs. (\ref{Psi}) do not mean the electric and magnetic fields.
We will see in the next section that the physics underlying the nonlocality of the photon spin is the nonlocality of the photon itself in the sense that such a wavefunction does not mean the probability amplitude for the position of the photon.

\section{No Probability Density Exists for the Position of Photons}

To this end, we rewrite the Dirac-like equation (\ref{DLE}) as
\begin{equation}\label{DLE-4D}
    \Gamma_0 \frac{\partial \Psi}{\partial x_0}
   +\Gamma_i \frac{\partial \Psi}{\partial x_i}=0.
\end{equation}
Its Hermitian conjugate is
\begin{equation*}
    -\frac{\partial \Psi^\dag}{\partial x_0} \Gamma_0
    +\frac{\partial \Psi^\dag}{\partial x_i} \Gamma_i =0.
\end{equation*}
Multiplying this equation by $\Gamma_0$ on the right, we have
\begin{equation}\label{DLE-TC}
    \frac{\partial \bar{\Psi}}{\partial x_0} \Gamma_0
   +\frac{\partial \bar{\Psi}}{\partial x_i} \Gamma_i =0,
\end{equation}
where
$\bar{\Psi}=\Psi^\dag \Gamma_0$. Multiplying Eq. (\ref{DLE-4D}) by $\bar{\Psi}$ on the left and Eq. (\ref{DLE-TC}) by $\Psi$ on the right, and summing, we get
\begin{equation*}
    \frac{\partial}{\partial x_\mu} (\bar{\Psi} \Gamma_\mu \Psi)=0.
\end{equation*}
This is the continuity equation for the four-current,
\begin{equation*}
    j_\mu =i c \bar{\Psi} \Gamma_\mu \Psi.
\end{equation*}
The time component
$j_0=ic \Psi^\dag \Psi$,
which corresponds to the positive-definite entity
$j_0/ic=\Psi^\dag \Psi$,
thus defines a constant of motion \cite{Cori-S},
\begin{equation}\label{P}
    P=\int \Psi^\dag \Psi d^3 x
     =\frac{1}{2} \int (\mathbf{F}_u^\ast \cdot \mathbf{F}_u
                       +\mathbf{F}_l^\ast \cdot \mathbf{F}_l) d^3 x.
\end{equation}
It can reasonably be interpreted as the total probability of the photon. However, the RQC (\ref{RQC}) makes it impossible to interpret the integrand $\Psi^\dag \Psi$ as the probability density for the position of the photon.

We have seen that the role of the RQC (\ref{RQC}) is played in momentum representation by Eqs. (\ref{CE-k}). Substituting Eq. (\ref{FT}) into Eq. (\ref{P}) and considering Eqs. (\ref{CE-k}), we find
\begin{equation*}
    P=\int \mathbf{f}_u^\ast \cdot \mathbf{f}_u d^3 k
     =\int \mathbf{f}_l^\ast \cdot \mathbf{f}_l d^3 k.
\end{equation*}
With the help of Eq. (\ref{FT-f}), it is transformed back into
\begin{equation*}
    P=\int \mathbf{F}_u^\ast \cdot \mathbf{F}_u d^3 x
     =\int \mathbf{F}_l^\ast \cdot \mathbf{F}_l d^3 x
\end{equation*}
in position representation, which is expressed in terms of the wavefunction $\Psi$ as
\begin{equation*}%\label{P-DE}
   P=\int \Psi^\dag (1+\Gamma_0) \Psi d^3 x
    =\int \Psi^\dag (1-\Gamma_0) \Psi d^3 x.
\end{equation*}
Since neither $\Psi^\dag (1+\Gamma_0)\Psi$ nor $\Psi^\dag (1-\Gamma_0)\Psi$
is equal to $\Psi^\dag \Psi$,
it is concluded that there is no unique probability density for the position of the photon. As a result, the wavefunction $\Psi$ does not mean the probability amplitude for the position of the photon as Inagaki \cite{Inag} speculated.
This shows that the photon is nonlocal in position space \cite{Jauch-P, Amre, Pauli, Rose-S}.

From the foregoing discussion, we can come to the conclusion that the RQC (\ref{RQC}) conveys the nonlocality of the photon in such a way that the position-representation wavefunction does not mean the probability amplitude for the position of the photon even though the integral of its modulus squared over the whole position space gives the total probability.
It should be emphasized that the nonlocality of the photon in position space does not mean the absence of a position-representation wavefunction as was stated in the literature \cite{Akhi-B, Sipe}.

\section{Conclusions and Remarks}

To conclude, we derived from the Darwin equations (\ref{REoM}) the spin of the photon. In particular, we showed that the RQC (\ref{RQC}) imposed on the wavefunction has a decisive impact on the properties of the photon spin. On one hand, it determines the operator for the photon spin to be
$(\mathbf{\Omega} \cdot \mathbf{w}) \mathbf{w}$
in momentum representation, which coincides with the well-known conclusion that the spin of the photon is always oriented in its propagation direction. This in turn makes it impossible to introduce the notion of local density for the photon spin in position space. On the other hand, it allows to use the position-representation wavefunction to express the expectation value of the photon spin as integrals of different integrands over the whole position space. The key point here is that confined by the RQC, the position-representation wavefunction no longer conveys the probability density for the position of the photon.
In addition, the Lorentz covariance of the Darwin equations was proven. The nonlocal dependence of the position-representation wavefunction on the classical fields was also discussed.

It is well known \cite{Stra,Jack} that the classical electric and magnetic fields can be used to construct an antisymmetric 4-tensor, the field strength tensor. But, as Cook \cite{Cook} showed, it is impossible to use the wavefunction to construct a 4-tensor though the Darwin equations (\ref{REoM}) can be cast into the form of free-space Maxwell equations (\ref{ME}).
Nevertheless, what is really quantized in modern quantum electrodynamics \cite{Akhi-B, Cohen-DG} or quantum optics \cite{Scul-Z} is such a wavefunction as is demonstrated by the relations (\ref{CR}) as well as the negative-energy solution of the Darwin equations.
According to Barnett \cite{Barn14}, the wavefunction satisfying the Darwin equations is a six-component spinor. The properties of this spinor need further investigation.

We have seen that due to the RQC, the operator for the photon spin in momentum representation does not satisfy the canonical commutation relation.
Instead, it is commutative. As a result, the operator for the photon orbital angular momentum in momentum representation does not satisfy the canonical commutation relation either if the total angular momentum is to satisfy the canonical commutation relation \cite{van-N1, van-N2}.
The only explanation for this is that the RQC makes the photon position, represented by $\mathbf{x}=i \nabla_k$ in momentum representation, not satisfy \cite{Li16}
$[x_i, x_j]=0$.
Otherwise, this equation together with the commutation relations (\ref{CR-PandX}) would lead to the canonical commutation relation of the orbital angular momentum \cite{Saku}, $[L_i, L_j]=i \hbar \epsilon_{ijk} L_k$.
This is also a demonstration of the nonlocality of the photon in position space. It is worth investigating further how the RQC determines the commutation relation of the photon position, in association with the spinor properties of the position-representation wavefunction. That is beyond the scope of present paper.

\section*{Acknowledgments}

The author is indebted to Kang-Rui Liu for his helpful discussions. This work was supported in part by the program of Shanghai Municipal Science and Technology Commission under Grant 18ZR1415500.

\appendix
%\renewcommand{\appendixname}{Appendix~\Alph{section}}
%\begin{appendices}
%\renewcommand\theequation{\Alph{section}.\arabic{equation}}

\section{Lorentz Transformation for the Wavefunction}\label{LT}

For simplicity we consider a Lorentz transformation of velocity $v$ along the first axis,
\begin{equation}\label{SLT}
    x'_\mu = a_{\mu \nu} x_\nu,
\end{equation}
for which the transformation matrix assumes the form
\begin{equation*}
    (a_{\mu \nu})=\left(
                    \begin{array}{cccc}
                      \gamma         & -i \beta \gamma & 0 & 0 \\
                      i \beta \gamma & \gamma          & 0 & 0 \\
                      0              & 0               & 1 & 0 \\
                      0              & 0               & 0 & 1 \\
                    \end{array}
                  \right),
\end{equation*}
where $\beta=v/c$ and $\gamma=1/(1-\beta^2)^{1/2}$. We will show that under this transformation the wavefunction transforms as follows,
\begin{equation}\label{LT-WF}
    \Psi' (x'_\mu) =\Lambda \Psi (x_\mu),
\end{equation}
where
$\Lambda=\exp(-i  \chi \Gamma_0 \Gamma_1)$
and $\chi= \cosh^{-1} \gamma$.

To this end, let us first show that Eq. (\ref{LT-WF}) can be cast into the transformation formulae for the electric and magnetic fields satisfying free-space Maxwell's equation under the Lorentz transformation (\ref{SLT}).
Letting
$\Psi'=\frac{1}{\sqrt 2} \left(
                           \begin{array}{c}
                             \mathbf{F}'_u \\
                             \mathbf{F}'_v \\
                           \end{array}
                         \right)
$ and noting that
\begin{equation*}
    \Lambda=1-i \Gamma_0 \Gamma_1 \sinh \chi -\Gamma_1^2 (1-\cosh \chi),
\end{equation*}
we can rewrite Eq. (\ref{LT-WF}) as
\begin{equation*}
    \left(
      \begin{array}{c}
        \mathbf{F}'_u \\
        \mathbf{F}'_v \\
      \end{array}
    \right)
   =\left(
      \begin{array}{cc}
        1-\Sigma_1^2 (1-\cosh \chi) & -i \Sigma_1 \sinh \chi \\
        i \Sigma_1 \sinh \chi & 1-\Sigma_1^2 (1-\cosh \chi)  \\
      \end{array}
    \right)
    \left(
      \begin{array}{c}
        \mathbf{F}_u \\
        \mathbf{F}_v \\
      \end{array}
    \right).
\end{equation*}
The components of the transformed wavefunction are therefore related to those of the original wavefunction via the following formulae,
\begin{subequations}\label{LT-1}
\begin{align}
  F'_{u1} & = F_{u1},     \\
  F'_{u2} & = F_{u2} \cosh \chi -F_{v3} \sinh \chi,     \\
  F'_{u3} & = F_{u3} \cosh \chi +F_{v2} \sinh \chi,
\end{align}
\end{subequations}
and
\begin{subequations}\label{LT-2}
\begin{align}
  F'_{v1} & = F_{v1},     \\
  F'_{v2} & = F_{v2} \cosh \chi +F_{u3} \sinh \chi,     \\
  F'_{v3} & = F_{v3} \cosh \chi -F_{u2} \sinh \chi.
\end{align}
\end{subequations}
They are the same as the transformation formulae \cite{Jack} for the electric and magnetic fields that satisfy free-space Maxwell's equations under the Lorentz transformation (\ref{SLT}). This indicates that Eq. (\ref{LT-WF}) is the transformation law for the photon wavefunction under the Lorentz transformation (\ref{SLT}). The invariance of Darwin's equations (\ref{REoM}) under the Lorentz transformations (\ref{SLT})-(\ref{LT-WF}) is shown below.

\section{Lorentz Covariance of Darwin's Equations}\label{LC}

Suppose that in the primed system we have the Darwin's equations,
\begin{subequations}
\begin{align}
  \Gamma_\mu p'_\mu \Psi' & =0,                                \label{DLE'} \\
  (\mathbf{\Gamma} \cdot \mathbf{p}')^2 \Psi' & =p'^2 \Psi'.   \label{RQC'}
\end{align}
\end{subequations}
Multiplying the Dirac-like equation (\ref{DLE'}) by $\Gamma_0$ on the left, we have
\begin{equation*}
    p'_0 \Psi' +(\Gamma_0 \mathbf{\Gamma} \cdot \mathbf{p}') \Psi' =0.
\end{equation*}
Upon substituting Eq. (\ref{LT-WF}), it becomes
\begin{equation*}
    p'_0 \Lambda \Psi +(\Gamma_0 \mathbf{\Gamma} \cdot \mathbf{p}') \Lambda \Psi =0,
\end{equation*}
which is equivalent to
\begin{equation*}
    p'_0 \Psi +\Lambda^{-1} (\Gamma_0 \mathbf{\Gamma} \cdot \mathbf{p}') \Lambda \Psi =0.
\end{equation*}
Noticing that
\begin{eqnarray*}
  \Lambda^{-1} \Gamma_0 \Gamma_1 \Lambda &=& \Gamma_0 \Gamma_1, \\
  \Lambda^{-1} \Gamma_0 \Gamma_2 \Lambda &=& \gamma \Gamma_0 \Gamma_2
  -i \beta \gamma (\Gamma_1 \Gamma_2 -\Gamma_2 \Gamma_1), \\
  \Lambda^{-1} \Gamma_0 \Gamma_3 \Lambda &=& \gamma \Gamma_0 \Gamma_3
  -i \beta \gamma (\Gamma_1 \Gamma_3 -\Gamma_3 \Gamma_1),
\end{eqnarray*}
and considering
$p'_\mu =a_{\mu \nu} p_\nu$, we find after lengthy but straightforward algebra,
\begin{equation}\label{IL}
    [(\Gamma_0-i \beta \Gamma_1)\Gamma_\mu p_\mu
    -i \beta (p_1 -\mathbf{\Gamma} \cdot \mathbf{p} \Gamma_1)] \Psi=0.
\end{equation}
It is noted that the second term of this equation on the left reads
\begin{equation*}
    (p_1 -\mathbf{\Gamma} \cdot \mathbf{p} \Gamma_1) \Psi
   =\left(
      \begin{array}{c}
        (p_1-\mathbf{\Sigma \cdot \mathbf{p}} \Sigma_1) \mathbf{F}_u \\
        (p_1-\mathbf{\Sigma \cdot \mathbf{p}} \Sigma_1) \mathbf{F}_v \\
      \end{array}
    \right),
\end{equation*}
which becomes
\begin{equation}\label{CC}
    (p_1 -\mathbf{\Gamma} \cdot \mathbf{p} \Gamma_1) \Psi
   =\left(
      \begin{array}{c}
        (\mathbf{p} \cdot \mathbf{F}_u) \mathbf{e}_1 \\
        (\mathbf{p} \cdot \mathbf{F}_v) \mathbf{e}_1 \\
      \end{array}
    \right)
\end{equation}
in accordance with relation
$(\mathbf{\Sigma} \cdot \mathbf{a}) \mathbf{b}= i (\mathbf{a} \times \mathbf{b})$.

On the other hand, as is discussed in Section \ref{DEs}, the RQC (\ref{RQC'}) in the primed system means
\begin{equation*}
    \nabla' \cdot \mathbf{F}'_u=\frac{\partial}{\partial x'_i} F'_{ui}=0, \quad
    \nabla' \cdot \mathbf{F}'_v=\frac{\partial}{\partial x'_i} F'_{vi}=0.
\end{equation*}
Upon using Lorentz transformation (\ref{SLT}) and transformation formulae (\ref{LT-1})-(\ref{LT-2}), it is not difficult to show that
\begin{equation*}
    \nabla' \cdot \mathbf{F}'_u=\gamma \nabla \cdot \mathbf{F}_u, \quad
    \nabla' \cdot \mathbf{F}'_v=\gamma \nabla \cdot \mathbf{F}_v.
\end{equation*}
As a result, we must have
\begin{equation}\label{ZD}
    \nabla \cdot \mathbf{F}_u=0, \quad \nabla \cdot \mathbf{F}_v=0,
\end{equation}
which means the RQC
\begin{equation*}
    (\mathbf{\Gamma} \cdot \mathbf{p})^2 \Psi =p^2 \Psi
\end{equation*}
in the unprimed system. This shows that the RQC (\ref{RQC}) is invariant under the Lorentz transformations (\ref{SLT})-(\ref{LT-WF}). With the help of Eqs. (\ref{ZD}), Eq. (\ref{CC}) reduces to
$(p_1 -\mathbf{\Gamma} \cdot \mathbf{p} \Gamma_1) \Psi=0$
and hence Eq. (\ref{IL}) becomes
\begin{equation*}
    (\Gamma_0-i \beta \Gamma_1)\Gamma_\mu p_\mu \Psi=0.
\end{equation*}
Because the matrix $\Gamma_0-i \beta \Gamma_1$ is  invertible, we finally get for the Dirac-like equation in the unprimed system,
\begin{equation*}
    \Gamma_\mu p_\mu \Psi=0.
\end{equation*}
In a word, the Dirac-like equation (\ref{DLE}) is invariant under the Lorentz transformations (\ref{SLT})-(\ref{LT-WF}).

From the above discussions we have a clear look at why the RQC (\ref{RQC}) is necessary for the Dirac-like equation (\ref{DLE}) to be Lorentz covariant.

%\end{appendices}

\end{document}